\documentclass[12pt]{spieman}
\usepackage{amsmath,amsfonts,amssymb}
\usepackage{graphicx}
\usepackage{setspace}
\usepackage{tocloft}

\usepackage{upgreek} %for upright Greek letters
\usepackage{booktabs} 
\usepackage[caption=false]{subfig}

\usepackage{xcolor} %for changing color of updated text

\newcommand{\ka}{K$\upalpha$}
\newcommand{\kb}{K$\upbeta$}
\newcommand{\microns}{$\upmu$m}

\title{Measuring the Soft X-Ray Quantum Efficiency of a Hybrid CMOS Detector}

\author[a,*]{Joseph M. Colosimo}
\author[a]{Abraham D. Falcone}
\author[a]{Mitchell Wages}
\author[a,b]{Samuel V. Hull}
\author[a]{Daniel M. LaRocca}
\author[a]{David N. Burrows}
\author[a]{Cole R. Armstrong}
\author[a]{Gooderham McCormick}
\author[a]{Mitchell Range}
\author[a]{Fredric Hancock}

\affil[a]{Pennsylvania State University, Department of Astronomy and Astrophysics, University Park, Pennsylvania, United States, 16802}
\affil[b]{Goddard Space Flight Center, X-Ray Astrophysics Laboratory, 8800 Greenbelt Rd, Greenbelt, Maryland, United States, 20771}

\graphicspath{{./}{Figures/}}

\cftpagenumbersoff{figure}
\cftpagenumbersoff{table} 
\begin{document} 
\maketitle

\begin{abstract}
Next-generation X-ray observatories, such as the Lynx X-ray Observatory Mission Concept or other similar concepts in the coming decade, will require detectors with high quantum efficiency (QE) across the soft X-ray band to observe the faint objects that drive their mission science objectives.
Hybrid CMOS Detectors (HCDs), a form of active-pixel sensor, are promising candidates for use on these missions because of their fast read-out, low power consumption, and intrinsic radiation hardness.
In this work, we present QE measurements of a Teledyne H2RG HCD, performed using a gas-flow proportional counter as a reference detector. 
We find that this detector achieves high QE across the soft X-ray band, with an effective QE of $94.6 \pm 1.1 \%$ at the Mn \ka{}/\kb{} energies (5.90/6.49 keV), $98.3 \pm 1.9 \%$ at the Al \ka{} energy (1.49 keV), $85.6 \pm 2.8 \%$ at the O \ka{} energy (0.52 keV), and $61.3 \pm 1.1 \%$ at the C \ka{} energy (0.28 keV). 
These values are in good agreement with our model, based on the absorption of detector layers.
We find similar results in a more restrictive analysis considering only high-quality events, with only somewhat reduced QE at lower energies.
\end{abstract}

\keywords{Hybrid CMOS, Soft X-Ray, Quantum Efficiency, Imaging Sensors, Detectors, Silicon}

{\noindent \footnotesize\textbf{*}Joseph M. Colosimo,  \linkable{jcolosimo@psu.edu} }

\section{Introduction}
\label{sect:intro} 

Next-generation X-ray observatories, potentially with capabilities similar to those of the proposed Lynx X-Ray Observatory \cite{Lynx19, Gaskin19}, will provide unprecedented observations of the early universe.
These missions will require large effective areas to enable detections of distant, faint sources. 
Detectors with capabilities surpassing those of the charge-coupled devices (CCDs) used in the focal planes of current missions will be required for future telescopes.
The detectors used in these observatories will need high quantum efficiency (QE), the fraction of incident X-rays individually detected, in order to fully utilize the large collecting areas provided by the mirror assemblies and maximize sensitivity to faint sources.
Future missions also require detectors with fast read-out in order to avoid pile-up, an effect in which multiple X-rays land in the same pixel during a single frame, when observing bright sources.
Without a major increase in the detector frame rate, pile-up would be exacerbated by the greater effective areas of these missions. 

\begin{figure}[bt]
    \centering
    \includegraphics[width=0.5\textwidth]{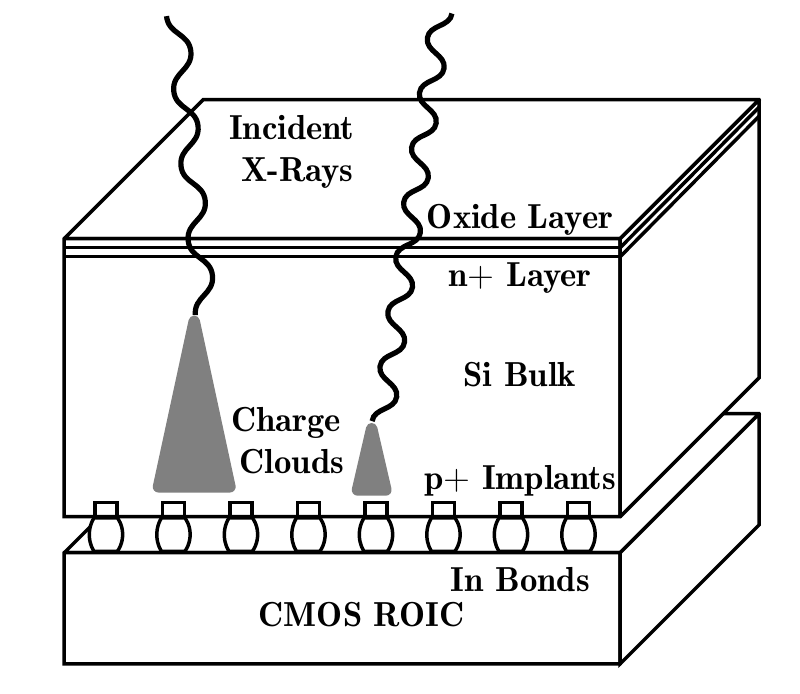}
    \caption{Diagram detailing the structure of silicon HCDs. In these detectors, X-rays absorbed in the absorber layer create a charge cloud which is collected and transferred to the CMOS ROIC to be measured and read out. The energy of an X-ray will impact its absorption depth and the size of its charge cloud. Adapted from Ref.~\citenum{Bongiorno15}.}
    \label{fig:detector_diagram}
\end{figure}

Hybrid CMOS Detectors (HCDs) are promising candidates for use in these future X-ray observatories.
HCDs are composed of a silicon absorber layer bonded to a complementary metal–oxide–\allowbreak{}semiconductor (CMOS) read-out integrated circuit (ROIC). The absorber layer produces photoelectrons when absorbing an X-ray and the ROIC reads the amount of charge generated. 
Each pixel in an HCD contains its own ROIC, removing the need to transfer charge through other pixels. This provides significantly faster readout, intrinsic radiation hardness, and lower power demands compared to the CCDs on current missions.
This hybridized structure, shown in Fig.~\ref{fig:detector_diagram}, allows for independent optimization of the ROIC and absorber layers during fabrication.
HCDs provide high QE across the soft X-ray band, due to their thin inactive surface layers and large depletion depths (100 \microns{} in current models) enabled by the use of high-resistivity silicon in the absorber layer. 

While the Penn State High-Energy Astrophysics Detector and Instrumentation Laboratory has previously measured a high QE on an H1RG HCD at the Mn \ka{}/\kb{} lines \cite{Bongiorno15}, further measurements are necessary to characterize the soft X-ray QE on other detectors and over a range of energies.
Characterization of the QE at low X-ray energies (0.2 - 1.5 keV) is crucial to detector development, as good low-energy performance is vital to the science goals of future missions.
For example, high QE at low energies would allow a mission like Lynx to more efficiently observe high-redshift quasars, whose power-law spectra have significant contribution in this band, and allow imaging of the halos of nearby galaxies, with low-energy thermal emission peaks \cite{Lynx19}.
In this work, we present measurements of the QE of an H2RG HCD at moderate and low energies. 
In particular, we report on the QE of this H2RG HCD at the Mn \ka{}/\kb{} (5.90/6.49 keV), Al \ka{} (1.49 keV), O \ka{} (0.52 keV), and C \ka{} (0.28 keV) characteristic X-ray emission lines.
Much of this work was presented at the 2021 SPIE Optics and Photonics meeting and is adapted from the corresponding proceedings\cite{Colosimo21}, with further expansions relative to that work.

\section{Quantum Efficiency Model}
\label{sec:model}

\begin{figure}[bt]
    \centering
    \includegraphics[width=0.6 \textwidth]{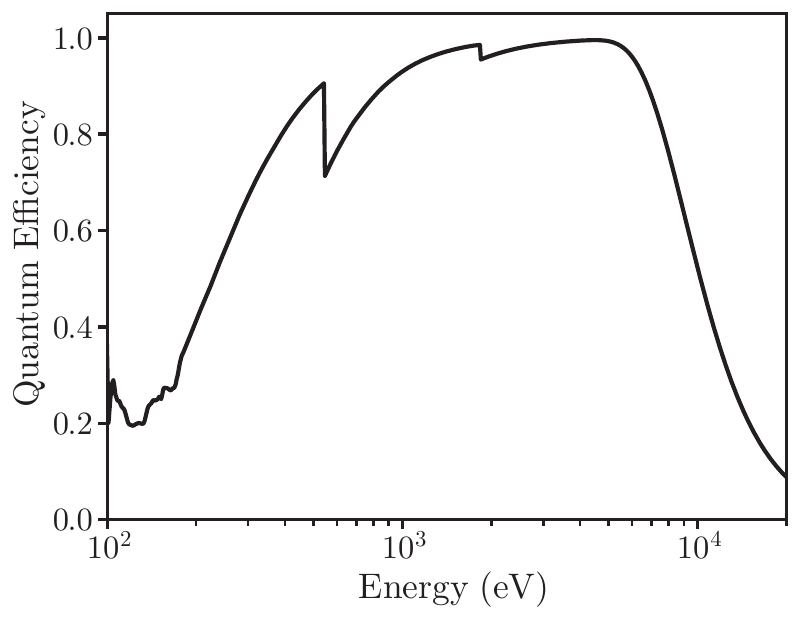}
    \caption{Estimate of the H2RG QE based on a 1D slab absorption model, using the nominal thicknesses of a 1028 \AA{} SiO$_2$ oxide layer and a 100 \microns{} absorber layer. The QE decreases at lower energies due to the increasing fraction of X-rays absorbed by the oxide layer and decreases again at high energies due to an increasing fraction of X-rays which are transmitted through the entire absorbing layer.}
    \label{fig:qe_model}
\end{figure}

We use a 1D slab absorption model to estimate the QE of HCDs. 
In this model, the QE is given by the fraction of X-rays transmitted through inactive surface layers and absorbed by the silicon absorber layer. 
This can be expressed in the following form:
\begin{equation}
    \textrm{QE}\,(E) = \left(1-e^{-\sigma_{\textrm{\tiny{Si}}}(E)\, n_{\textrm{\tiny{Si}}} t_{\textrm{\tiny{abs}}}}\right)\prod^{N}_{i}e^{-\sigma_{i}(E)\, n_{i} t_{i}}
    \label{eq:model}
\end{equation}
where $t_\textrm{abs}$ is the thickness of the absorbing layer (with the cross section and number density for silicon) and the product is over the atomic species in the inactive surface layer(s) with atomic cross sections $\sigma_i$, number densities $n_i$, and corresponding layer thicknesses $t_i$. 
The H2RG has a 100 $\upmu \rm{m}$ Si absorber layer and a single inactive surface layer composed of SiO$_2$ with a nominal thickness of 1028 \AA{} \cite{Bai21}.
In this work, we calculate transmission values using the absorption coefficients provided by Ref.~\citenum{Henke93}. The resulting H2RG QE model is shown in Fig.~\ref{fig:qe_model}.

At energies below 5 keV, few X-rays are transmitted through the Si absorbing layer, and the QE is primarily determined by the transmittance of the inactive surface layer. 
Above 5 keV, almost all X-rays are transmitted through the inactive surface layer, and the QE is primarily determined by the fraction of X-rays absorbed by the Si absorbing layer.  
While the simple slab absorption model has proven accurate in past HCD QE measurements \cite{Bongiorno15, Prieskorn14, Kenter05}, effects ignored by this simple 1D model, such as charge traps near the Si/SiO$_2$ interface, may cause deviations from model values. 
This model does not consider the detector response to absorbed X-rays, which will impact the QE achievable by a detector.
Charge diffusion and excessive read noise may cause a sub-optimal detector response to absorbed events, preventing their energies from being properly identified \cite{Miller18}.
These various effects can combine to complicate the response of pixelated silicon detectors, making measurements essential to truly understand their behaviour.
QE measurements over a range of energies are therefore necessary to fully characterize the performance of detectors, to determine their suitability for future missions, and to guide the development of future detectors.

\section{Experimental Methods}
\label{sect:methods}

\begin{figure}[bt]
    \centering
    \includegraphics[width= 6.5 in]{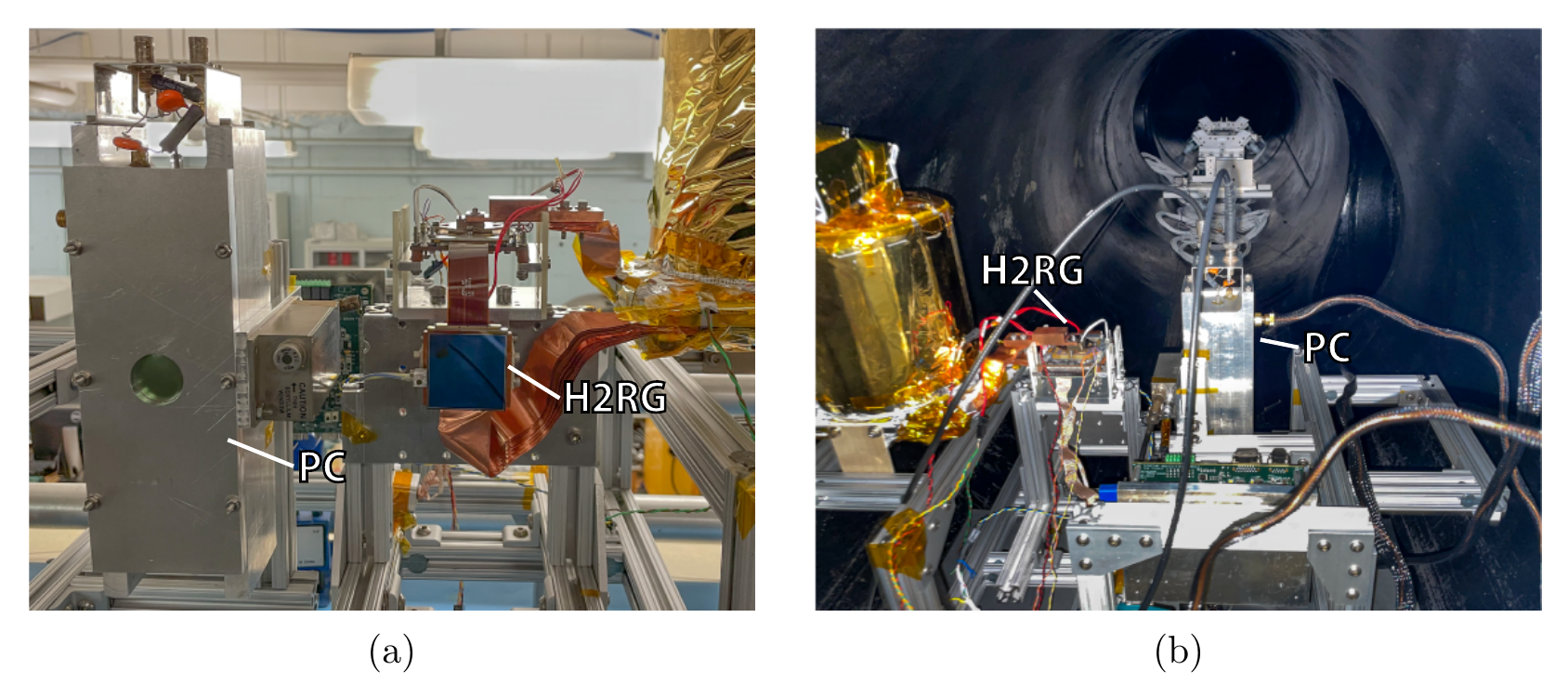}
    \caption{The front of the detector mounting structure shown outside the vacuum chamber (a), with the PC (left) and the H2RG (right), and the back of the same structure shown in place for the QE measurements inside the beamline (b). A Si PIN diode detector was also mounted between these detectors but was not used during QE measurements.}
    \label{fig:detectors}
\end{figure}

QE measurements require a separate, well-characterized reference detector to determine the incident X-ray flux.
We used a gas-flow proportional counter (PC) as our reference detector because of its known QE and high response across the soft X-ray band. 
During the experiment, both the H2RG and PC were simultaneously exposed at a similar distance to the same source.
The flux measured by the H2RG was compared to the flux measured by the reference detector to determine the QE.
The QE measurements were conducted in the 47-meter X-ray beamline operated by the Penn State High-Energy Astrophysics Detector and Instrumentation Laboratory.
The mounting structure containing both detectors can be seen in and outside of the beamline in Fig.~\ref{fig:detectors}.
The large distance between the source and detectors as well as the simultaneous data collection minimized the difference in flux incident upon both detectors.
The beamline is maintained at high vacuum ($<10^{-5}$ mbar) to minimize attenuation of X-rays and avoid the condensation of water vapor onto the HCD (which is cooled during operation).

\subsection{H2RG}
\label{sec:h2rg}

We conducted our QE measurements on a modified X-ray H2RG fabricated by Teledyne Imaging Sensors.
This detector uses a standard H2RG ROIC, an array of $2048\times2048$ pixels with an 18 $\upmu$m pitch, bonded to an array of $1024\times1024$ 36 $\upmu$m pitch Si absorber cells, with only one out of every four ROIC pixels connected to the absorber layer. 
This design was selected to increase the distance between pixels in order to reduce the effect of interpixel capacitance, a form of crosstalk which decreased the energy resolution of older-generation H1RG HCDs \cite{Falcone12}.   
While extensive characterization measurements have been taken for this detector \cite{Prieskorn13}, and it even has flight heritage on the Water Recovery X-ray Rocket mission \cite{Wages19}, no QE measurements had been made on the X-ray H2RG prior to this work.

X-ray hybrid CMOS detectors must be cooled during operation to minimize dark current and instrumental background noise. 
The detector was cooled using a copper strap thermally connecting the device to an internal liquid nitrogen dewar and the temperature was regulated using a heater. 
When possible, the detector was maintained at 160 K during measurements. 
This temperature is slightly higher than the optimal operating temperature of $\lesssim 150$ K, leading to increased noise and reduced energy resolution; however, we found that operating below 160 K allowed for deposition of an ice layer on the detector surface, which reduced the measured QE. 
This effect was observed as a gradual degradation of the QE with time.
We installed a residual gas analyzer (RGA) to measure the partial pressure of water vapor in the chamber and confirmed that it was lower than the deposition pressure at 160 K during our QE measurements \cite{Wagner11}. 
During the lowest-energy measurement (C \ka{}), the detector was cooled to 140 K in order to suppress the instrument noise peak sufficiently to obtain satisfactory low-energy data. 
The deposition pressure of water vapor at this temperature is less than the partial pressure which was measured by the RGA during these measurements, allowing a layer of ice to form on the detector surface.

\subsection{Proportional Counter}
\label{sec:pc}

The PC and H2RG were mounted side-by-side and equidistant from the midline of the chamber, as shown in Fig.~\ref{fig:detectors}. 
This arrangement provides roughly equal flux incident on both detectors. 
P-10 gas (
90\% argon and 10\% methane, with a tolerance of $2\%$) was flowed through the PC at a rate of $\sim200$ cc/min.
The temperature of the gas was measured to be $290\pm5$ K.
The pressure of the gas was recorded for each measurement (as it was affected by variations in atmospheric pressure), with a typical value of  1030 mbar and an estimated uncertainty of 2.7 mbar.
The gas is separated from vacuum by a thin window installed in the aperture of the detector. 
The window was manufactured by Luxel Corporation for high transmittance of low-energy X-rays and was constructed from aluminized polyimide, supported by a stainless steel mesh. 

In order for the PC to act as a reference detector and accurately measure the absolute flux produced by the source, its QE must be well-characterized.
PCs are commonly used as reference detectors because their QE can be reliably determined through measurements of gas properties and window transmittance.
We calculated the QE using a model similar to the HCD model described in \S\ref{sec:model}, where the PC QE is given by the fraction of X-rays transmitted through the window which are then absorbed by the P-10 gas. 
The gas absorption was calculated based on the pressure, temperature, and composition of the P-10 gas, while the transmittance of the window in the PC aperture was measured to ensure accuracy. 

Measurements of the window transmittance were necessary to obtain an accurate absolute calibration of the PC, as slight variations in the thickness of the window layers can cause significant changes in its transmittance. 
In the low-energy band (0.2 - 1.5 keV), almost all X-rays transmitted by the window are absorbed by the P-10 gas, so the QE of the detector is primarily a function of the window transmittance. 
Calibration measurements are also important at high energies, as the stainless steel mesh used to support the window acts as a binary mask, blocking X-rays at all energies throughout the soft X-ray band.

\subsubsection{Proportional Counter Calibration}
\label{sec:pc_calibration}

We measured the transmittance of the PC window using a secondary window which was also manufactured by Luxel with the same specifications as the primary window. 
The secondary window was installed in the aperture of the PC during the calibration measurements. 
We then mounted the primary window directly in front of the secondary window and compared the count rates when the primary window was present and when it was removed.
The ratio of the observed count rates in these measurements gives the transmittance of the window at a given energy. 

\begin{figure}[bt]
    \centering
    \includegraphics[width=0.6 \textwidth]{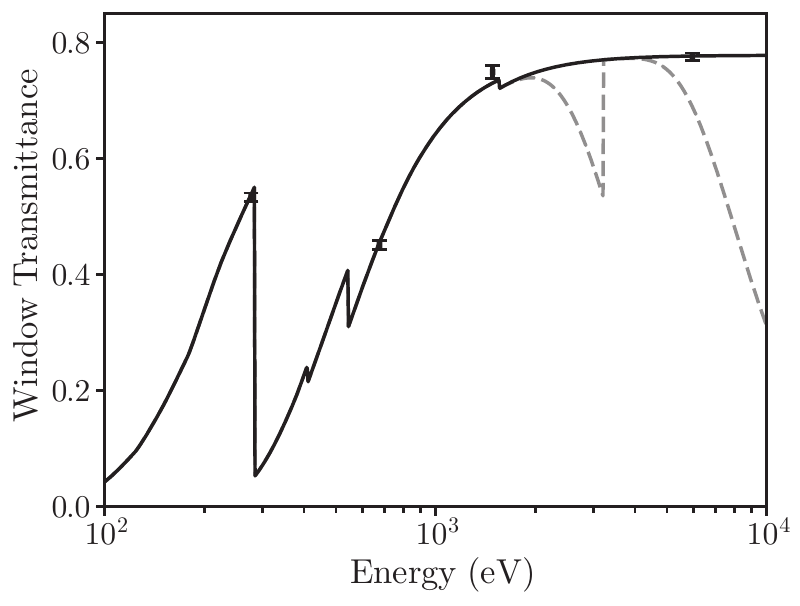}
    \caption{Measurements of PC window transmittance at four energies shown on top of the fitted transmittance model (solid black line). The modeled PC QE is shown by the dashed gray line. Note that the QE and window transmittance models only diverge at higher energies, where a significant fraction of X-rays are transmitted through the gas undetected.}
    \label{fig:pc_window}
\end{figure}

We used radioactive sources because these provide a known flux, decaying at a known rate.
An Fe-55 source was used for a measurement of the effective window transmittance at the Mn \ka{}/\kb{} energies. 
We made transmittance measurements at lower energies using targets fluoresced by Polonium-210 alpha-particle sources. 
Measurements were made at the Al \ka{}, F \ka{} (0.68 keV), and C \ka{} lines using fluoresced aluminum, Teflon, and polyethylene targets respectively.
These measurements were directly used in the calculation of PC QE for the Mn \ka{}/\kb{}, Al \ka{}, and C \ka{} H2RG QE calculations.

We did not make a measurement at the O \ka{} energy; however, we were able to fit a model to calculate the transmittance at this energy. 
In our model, the window layer thicknesses were fixed at the values provided by Luxel (listed in Table~\ref{tab:pc_window}), and only the mesh transmittance was allowed to vary.
A mesh transmittance of $0.780\pm0.005$ provided the best fit to our measurements.
We found that this fitted model agrees quite well with our measurements and used it to determine the window transmittance at the O \ka{} energy. 
This fit is shown in Fig.~\ref{fig:pc_window}, along with the measured transmittance values and the modeled PC QE.

\begin{table}[bt]
    \caption{Properties of the various layers of the PC window. Nominal thicknesses and estimates of the uncertainty for each layer were provided by Luxel \cite{Zeiger19}. These values were used to model the PC window transmittance and calculate the PC QE at the O \ka{} energy. The densities of the aluminum and carbon layers are assumed to be $90\%$ of their bulk densities because of the method of fabrication.}
    \begin{center}
    \begin{tabular}{ccc}
    
    \toprule
    \textbf{Window Layer} & \textbf{Thickness (nm)} & \textbf{Uncertainty (nm)} \\
    \midrule
    Carbon         &   18.0  & 0.9  \\
    Polyimide      &   440.0 & 22.0 \\
    Aluminum       &   19.0  & 0.95 \\
    Aluminum Oxide &   3.0   & 1.0  \\ 
    
    \bottomrule
    \end{tabular}
    \label{tab:pc_window}
    \end{center}
\end{table}

\subsection{X-Ray Sources}
\label{sec:sources}

We used two different sources of X-rays for our QE measurements.  A radioactive Fe-55 source produced X-rays at the Mn \ka{} and \kb{} energies, and a Manson electron-impact X-ray source produced lower-energy X-rays at the Al, O, and C \ka{} lines. These sources required careful calibration to ensure that both the energies and spatial distribution of the X-rays were well known. Both sources and their calibrations are described in this section.

\subsubsection{Fe-55 Source}
\label{sec:fe55}

We used an Fe-55 radioactive source to produce emission at 5.9 and 6.5 keV.
We mounted the Fe-55 source on the detector end of the beamline in order to produce a sufficient count rate. 
Unlike the Manson source, the Fe-55 source emits isotropically, so the incident flux is only a function of distance to the source.
The Fe-55 source is not exactly equidistant from both detectors ($2.450\pm0.010$ m from the H2RG and $2.455\pm0.010$ m from the PC), leading the flux incident on the H2RG to be $0.57\pm0.04\%$ greater than the flux incident on the PC.
The source was placed on an electronic linear stage so that it could be moved behind a shutter when not in use, allowing for dark or Manson measurements without opening the chamber. 

The PC does not have sufficient energy resolution to distinguish between the \ka{} and \kb{} peaks. Because the QE of the PC differs significantly at these energies (as seen in Fig.~\ref{fig:pc_window}), knowing the fraction of incident X-rays at each energy is essential to accurately determining the fraction of X-rays absorbed by the P-10 gas.
The intrinsic ratio of Mn \ka{} to \kb{} emission is 0.1195 \cite{scofield74}; however, the emitted line ratio is increased by the greater transmission of \kb{} X-rays through this Fe-55 source and its casing. 
We calculated the emitted \kb{} to \ka{} X-ray ratio to be $0.132\pm0.002$ based on the thicknesses of the Fe-55 source layer and Ni casing provided by the manufacturer. 

\subsubsection{Manson Source}
\label{sec:manson}

\begin{figure}[tb]
    \centering
    \includegraphics[width=0.6 \textwidth]{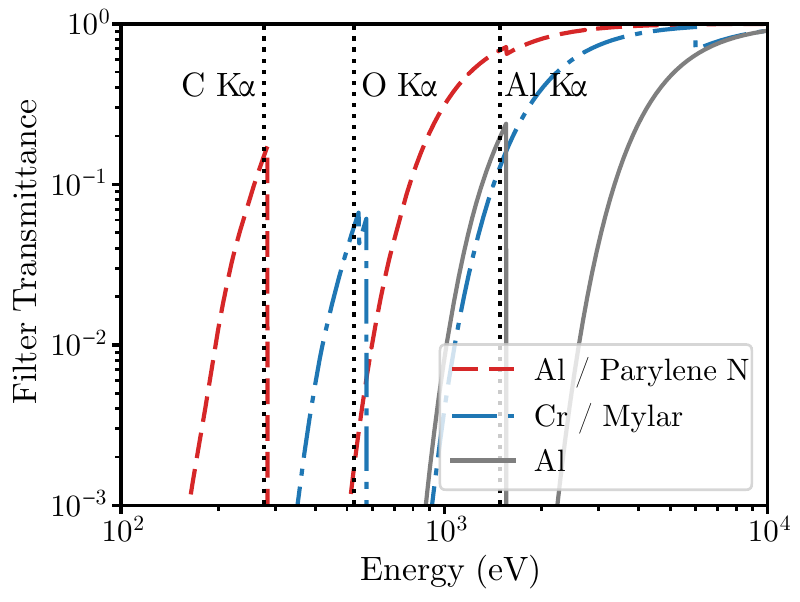}
    \caption{Model of filter transmittance as a function of energy for the filters used with the Manson source. The absorption edges make these filters extremely efficient at blocking continuum X-rays just above the characteristic energy.}
    \label{fig:filters}
\end{figure}

A Manson ultra-soft X-ray source, a type of electron-impact source, was used to produce X-rays for measurements at lower energies.
We used an aluminum anode in the Manson source for QE measurements at the Al \ka{} line  (1.49 keV), a heavily oxidized magnesium anode for measurements at the O \ka{} line (0.52 keV), and a graphite anode for measurements at the C \ka{} line (0.28 keV). 
In addition to characteristic lines, the Manson source also emits continuum X-rays. 
These X-rays must be suppressed in order to prevent X-rays at different energies from being included in the characteristic X-ray fluxes used in the QE measurements. 
We installed carefully-designed filters in the source end of the beamline to block continuum X-rays. 
These filters are also required to block optical light, produced by the Manson source, predominantly from the hot filament.
If not blocked, this light would be registered by the detector, adding noise to our measurements and potentially damaging the CMOS ROIC.
We used an aluminum filter, constructed from a piece of heavy-duty aluminum foil ($\sim 25$ \microns{}), for the Al \ka{} measurements.
The lower-energy measurements required more carefully designed filters made from thin foils and films, manufactured by Lebow Company, allowing for adequate transmittance of low-energy X-rays. 
The filter for the O \ka{} measurement was constructed from a 1-\microns{} Cr foil supported by a 0.5-\microns{} Mylar film, and the filter for the C \ka{} measurement was constructed from a 0.1-\microns{} Al foil supported by a 5-\microns{} parylene N film. 
Models of the transmittance for each filter are shown in Fig.~\ref{fig:filters}.
The filters do not completely eliminate the continuum emission; however, they provide sufficient separation between the characteristic X-ray and continuum peaks so that both can be modeled and the counts can be separated.

While the large distance between the Manson source and the detectors ($\sim46 \pm 0.1$ m) allowed for fairly uniform flux incident on the detectors, we found that it was neither perfectly uniform nor perfectly symmetric. 
The detectors were positioned on the mounting structure such that the distances from each detector to the source would be identical to a tolerance of 1 cm, much too small to explain the observed differences in incident flux. 
These spatial variations are instead thought to result from inherent properties of the source setup (e.g., variations in the filter thicknesses).
We obtained a calibration for each Manson source and filter configuration using a detector mounted on a linear stage. 
We measured the count rate at the position of both the H2RG and PC. 
These measurements were repeated to account for variability in the absolute flux produced by the Manson source. 
We found that the percentage of the flux incident upon the H2RG relative to that incident on the PC was $99.1\pm1.0\%$ in the  Al \ka{} configuration, $100.0\pm0.2\%$ in the O \ka{} configuration, and $99.3 \pm 0.4\%$ in the C \ka{} configuration.

\section{Analysis and Results}
\label{sec:analysis}

During data acquisition, both detectors measured the incident flux produced by a given source simultaneously. 
At each of the energies measured, we recorded data with the detectors exposed to the source then subsequently recorded background measurements with the source blocked. 
Each measurement lasted for a duration of approximately 10 minutes (with exact live times for the detectors differing slightly). 
In this section, we describe the methods used to analyze this data, discuss the QE calculation, and present the results of our measurements.

\subsection{H2RG Data Reduction}

During these measurements, we operated the H2RG in ``up-the-ramp'' mode, in which many frames were taken successively without clearing charge using non-destructive readout. 
The charge accumulated in a single frame was found by taking the difference in pixel values between consecutive frames. 
Row and channel noise were then removed from the image using a boxcar smoothing algorithm.
In order to identify X-ray events, we found pixels which had charge depositions that were local maxima and above a set ``event'' threshold.
If the charge in a surrounding pixel exceeded a ``split'' threshold, it was included in the total charge of the X-ray event.

We also used the split threshold to assign a grade to each event, based on the distribution of pixels included in the events.
This allows us to characterize the degree of charge spreading and enables discrimination between events corresponding to X-rays and those caused by cosmic rays or noise. 
The selection of allowed grades has a significant effect on the measured QE.
Including all events, regardless of grade, will give an idealized QE for a detector, while a flight-like analysis which restricts the allowed event grades may give a more realistic QE.
In this work, we investigate both approaches, measuring the QE when considering all event grades and when only considering events with high-quality grades.
For our more restrictive analysis, we select events with grades 0 -- 12 in the Swift XRT photon-counting mode grading scheme \cite{Burrows05}, allowing for the inclusion of 1 -- 4 pixel events with shapes expected for X-ray events. 

We use different event and split thresholds for each energy in order to optimally exclude noise and include charge in each analysis. 
These different values are shown in Table \ref{tab:thresholds} in terms of the measured detector noise ($\sigma_n$). 
We use high thresholds in the Mn \ka{}/\kb{} and Al \ka{} analyses in order to clearly separate pixels with charge from X-ray events from those with spurious values due to noise. 
The lower-energy O \ka{} and C \ka{} analyses require lower event thresholds, as higher values can miss a significant fraction of events and reduce the measured QE. 
We also reduce the split thresholds in order to more accurately sum charge for low-energy events.
Given these $2\sigma_n$ split thresholds, we expect a considerable number of X-ray events to include split pixels with high values due to random noise and not charge spreading, artificially skewing the energy distribution and giving some events low-quality grades.

\begin{table}[bt]
    \caption{Thresholds used for X-ray event identification and evaluation at different energies, given in terms of the detector noise ($\sigma_n$). Note that the noise value for the C \ka{} measurement is lower due to the colder detector operating temperature used during this measurement.}
    \begin{center}
    \begin{tabular}{lccc}
    \toprule
    \textbf{Emission Line} & \textbf{$\sigma_n$ (e$^-$)} & \textbf{Event Threshold ($\sigma_n$)} & \textbf{Split Threshold ($\sigma_n$)}\\
    \midrule
    Mn \ka{}/\kb{} & 12 & 10 & 5 \\
    Al \ka{}     & 12   & 8  & 4 \\ 
    O \ka{}      & 12   & 3 & 2 \\
    C \ka{}      & 6    & 4 & 2 \\
    \bottomrule
    \end{tabular}
    \label{tab:thresholds}
    \end{center}
\end{table}

Our H2RG is an engineering-grade detector, with a greater number of defects in the silicon lattice of the absorber layer compared to a science-grade sensor.
It has many charge traps which create areas with significant dark current and limited dynamic range.
We chose to exclude X-ray events from pixels in regions associated with these charge traps, as these pixels may have reduced sensitivity to incident X-rays.
A science-grade detector would have fewer defects, so this measurement better reflects the expected QE of the detectors which would be used on a mission.
We also excluded events from pixels too close to the edge of the detector, where edge effects reduce the detector sensitivity. 
After masking off these pixels, we are left with 85.7\% of the total detector area as active. 
The area of the detector was scaled by this value in calculations of the measured flux and the QE.

\subsection{Mn \texorpdfstring{\ka{}/\kb{}}{Ka/Kb} Measurement}

\begin{figure}[bt]
    \centering
    \includegraphics[width= 6.5 in]{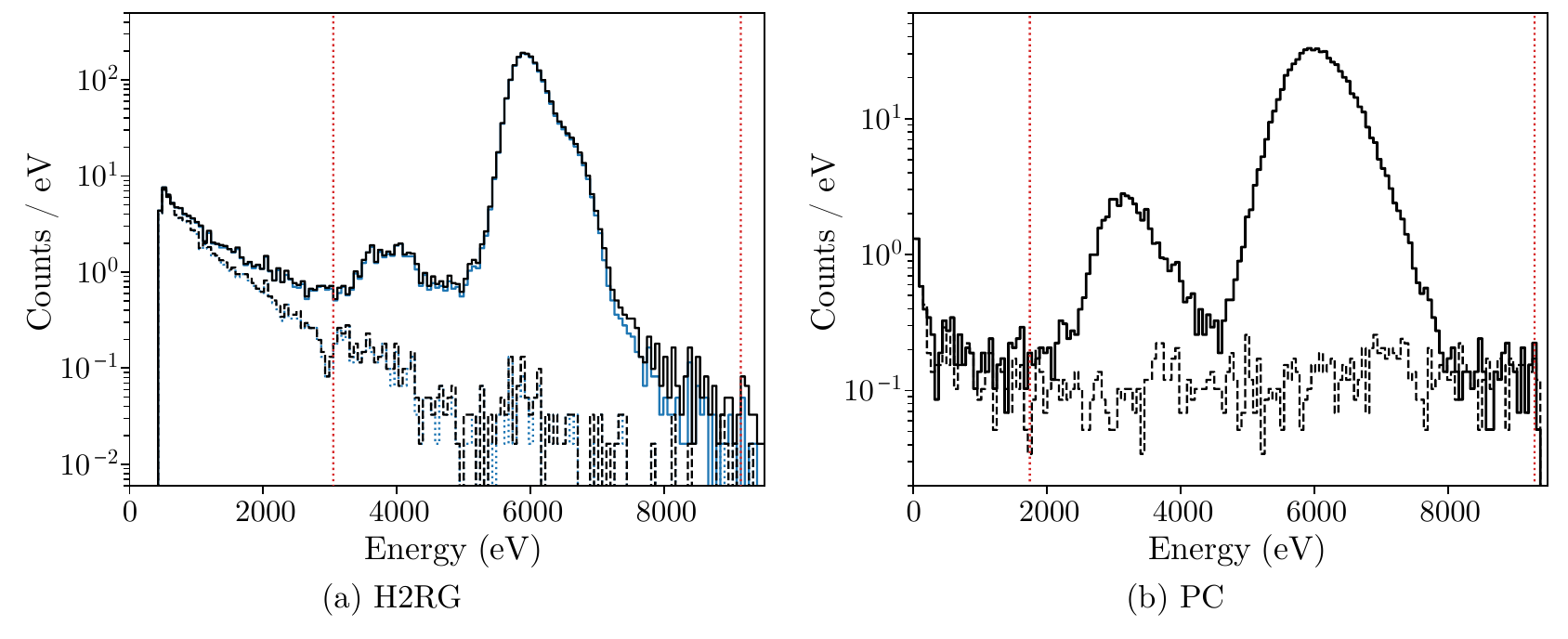}%
    \caption{H2RG and PC spectra from the Fe-55 QE measurements. Both spectra include combined Mn \ka{}/\kb{} peaks with escape peaks at lower energies. The (scaled) background spectra from the dark measurement are shown by the dashed lines. The bounds used to sum the counts and determine the fluxes are shown by the dotted red lines. The H2RG spectra with all event grades are shown in black, while the spectra with grades 0 -- 12 are shown in blue. Applying this grades cut did not significantly change the Fe-55 or background spectra, as shown by the nearly overlapping lines. The location of the Mn \ka{} peak from this spectrum was used to calculate the gain used in all of the plotted H2RG spectra in this work (PC gain varied between measurements and was fit separately in each measurement).}
    \label{fig:spec_Mn}
\end{figure}

Fig.~\ref{fig:spec_Mn} shows the H2RG and PC spectra from the Fe-55 QE measurement.
Both the H2RG analysis including all event grades and that including only grades 0 -- 12 are shown in this plot, revealing only a small reduction in the number of events due to the grades cut.
Due to the poor energy resolution of the PC and H2RG (when operated at 160 K), the Mn \ka{} and \kb{} peaks are blended together in the spectra from both detectors. 
For this reason, we make an effective QE measurement at the combined Mn \ka{} and \kb{} energies, giving the fraction of all incident Mn K X-rays detected.

We determine the count rates for both detectors by summing all counts between specified bounds. 
The sums of the counts in the background spectra within these bounds are subtracted from these count rates, removing the instrumental backgrounds and leaving only the count rates from real X-ray events. 
This is an appropriate method for measuring the count rate of Mn K X-rays, as no other sources of characteristic or continuum emission were present during this measurement.
Spectra from both detectors include escape peaks, as the Mn \ka{}/\kb{} X-rays are more energetic than the binding energies of the Ar and Si K shells. 
The escape peaks were included within the bounds of the sums because they correspond to real Mn K X-rays registered by the detectors whose energies can be accurately determined given a proper detector response matrix. 
The H2RG spectrum also includes substantial high- and low-energy tails, likely due to thresholding effects; however, as most counts from these features are included in our bounds, they should have little effect on the measured QE.

\subsection{Al \texorpdfstring{\ka{}}{Ka}, O \texorpdfstring{\ka{}}{Ka}, and C \texorpdfstring{\ka{}}{Ka} Measurements}

The Al \ka{}, O \ka{}, and C \ka{} spectra contain continuum emission from the Manson source in addition to the characteristic X-ray peaks. 
This continuum emission must be separated out to obtain accurate QE measurements.
For this reason, we chose to model and fit these spectra, allowing us to use only the model components corresponding to the characteristic emission in our QE analysis. 
We removed instrument background counts from each of the spectra before fitting by subtracting the binned counts of the dark spectra (scaled based on live time) from those of the measurement spectra. 
The models, described below, were fit using a least-squares algorithm.
The total number of counts was found by integrating the area under the fitted curves of the model components corresponding to characteristic X-ray peaks,
with lower integration bounds of zero and upper bounds set individually to fully assess the flux for each measurement.

\begin{figure}[b]
    \centering
    \includegraphics[width= 6.5 in]{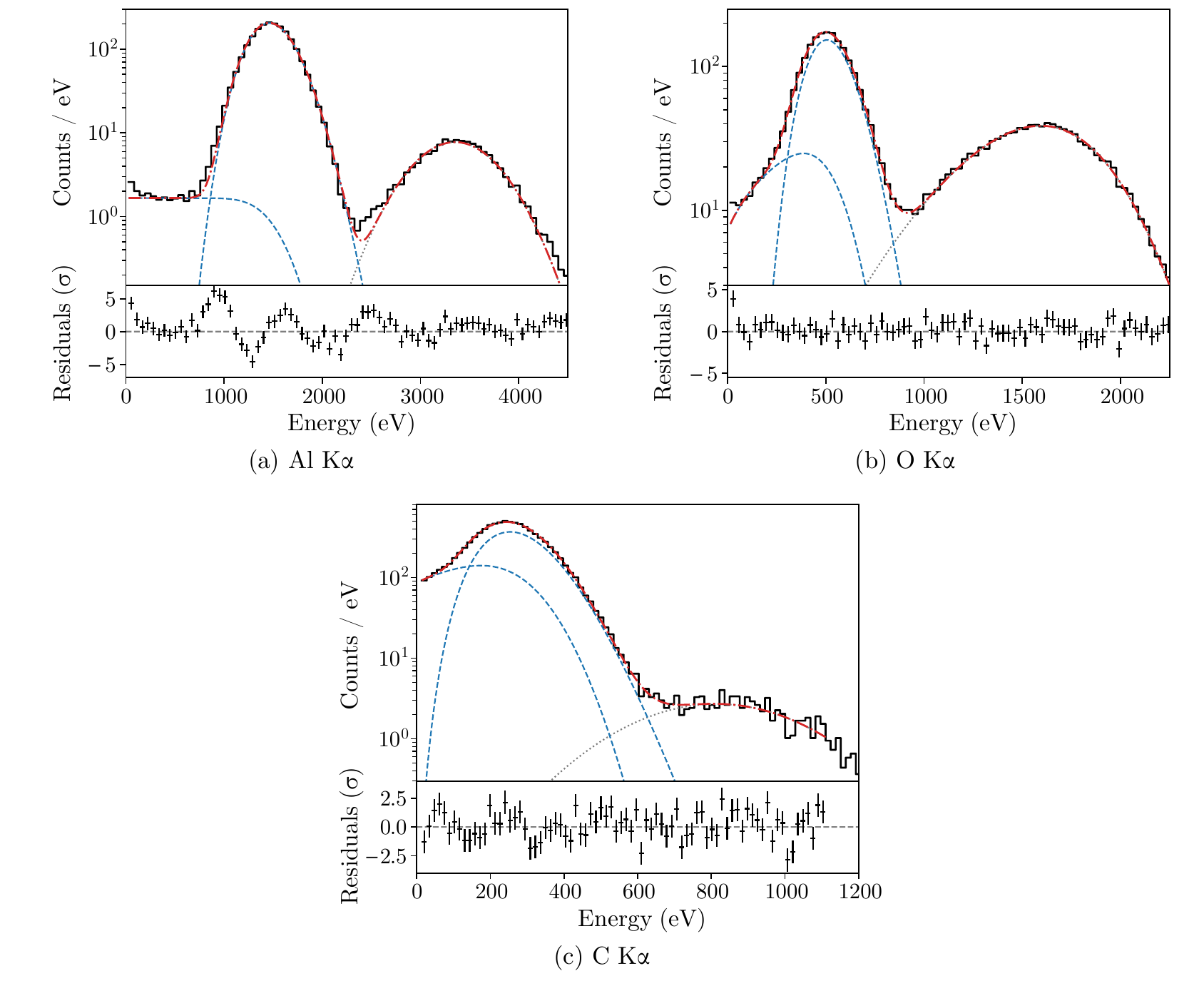}
    \caption{PC spectra for the Al, O, and C \ka{} QE measurements along with the spectral models (red dash-dotted lines) and residuals plotted below. The model components included in the PC characteristic flux (Prescott function and low-energy shelf) are shown by the blue dashed lines while the excluded continuum flux is shown by the gray dotted lines.}
    \label{fig:pc_spec}
\end{figure}

\subsubsection{PC Models}

The PC spectra, fitted models, and model components are shown in  Fig.~\ref{fig:pc_spec}.
The primary features of these spectra are skewed characteristic peaks with low-energy shelves along with a peak from continuum emission at higher energies.
The low-energy shelves are the results of events with partial charge collection, where a fraction of the charge recombines before amplification and collection. 
We modeled the characteristic peaks and low-energy shelves using the distributions described in Ref.~\citenum{Auerhammer98}. 
The characteristic peaks were modeled as Prescott functions \cite{Prescott63}:
\begin{equation}
    P(x) = \frac{P_0\;x_0^{1/4} }{x^{3/4}\sqrt{4\pi Q}} \exp{\left[-\left(\sqrt{x}-\sqrt{x_0}\right)^2/Q\right]},
    \label{eq:prescott}
\end{equation}
where $x_0$ is a parameter which defines the location of the peak, $Q$ defines its width, and $P_0$ defines its amplitude. 
The low-energy shelves were modeled as the following: 
\begin{equation}
    S(x) = (a+bx) \, \left(1-\textrm{erf}\left[\frac{x-x_0}{2\sqrt{2} \, x_0 \, Q}\right]\right)
    \label{eq:shelf}
\end{equation}
where $x_0$ and $Q$ are parameters from the Prescott function and $a$ and $b$ are parameters defining the amplitude and slope of the shelf. 
In the Al \ka{} spectrum, the slope of the shelf ($b$) is set to zero because this parameter is not needed to achieve a good fit. 
The higher-energy continuum peaks are modeled as skew-normal curves in the Al and O \ka{} spectra and as a Gaussian curve in the C \ka{} spectrum.
The high-energy peak in the O \ka{} spectrum also includes Mg \ka{} counts; however, a single skew-normal curve was sufficient to model both the continuum and Mg \ka{} components due to the limited spectral resolution of the PC.

\begin{figure}[b]
    \centering
    \includegraphics[width= 6.5 in]{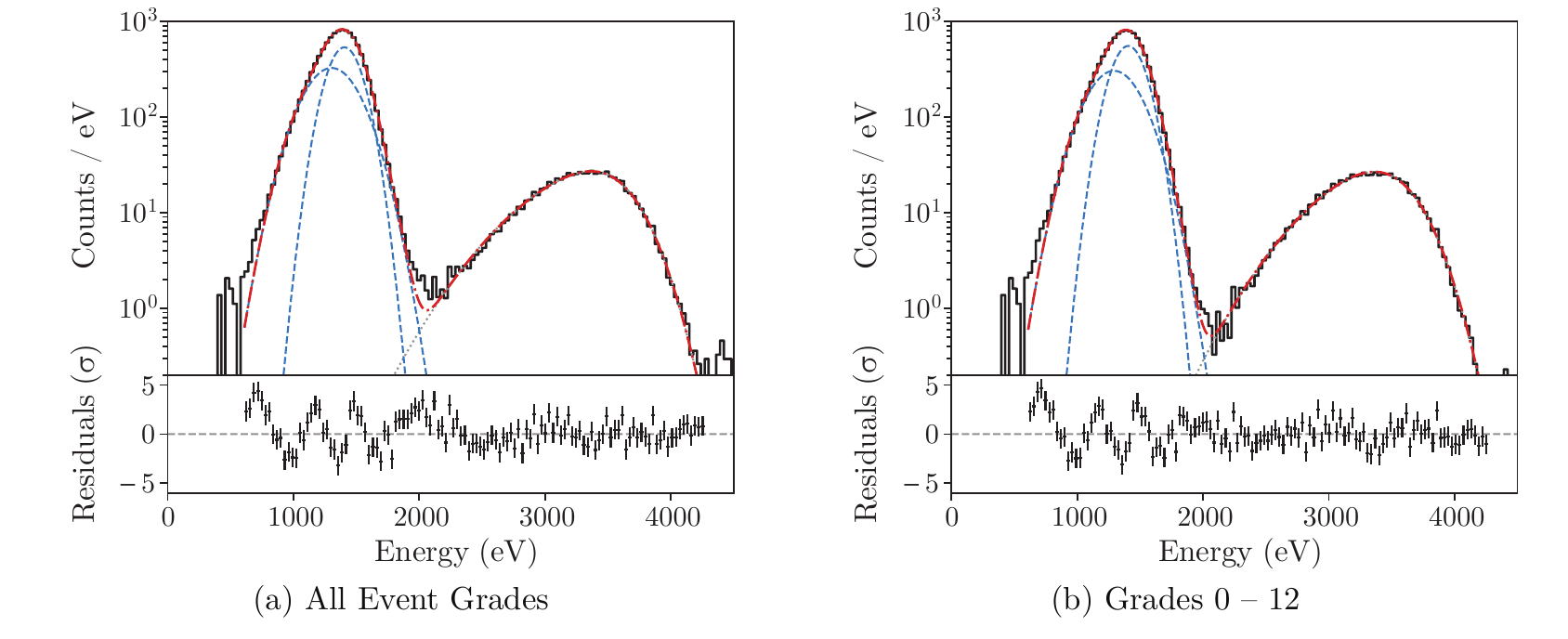}
    \caption{The H2RG spectra from the Al \ka{} measurement along with the models and residuals. The full model is shown as the red dash-dotted line, while the characteristic model components are shown by the  blue dashed lines and the excluded continuum model is shown by the gray dotted line. Both grades filters result in similar spectra, with slightly better differentiation between the characteristic and continuum peaks in the graded spectrum.}
    \label{fig:spec_Al}
\end{figure}

\break

\subsubsection{H2RG Models}

We fit two different models for each of the H2RG spectra, one for the spectrum including all events and another for those with only grades 0 -- 12.
We run two separate analyses for the different grades filters, fitting a model separately for each spectrum. 
The Al \ka{} spectra for both grades filters are shown in Fig.~\ref{fig:spec_Al} along with the fitted models and model components.
In both analyses, the characteristic peak is modeled as a sum of two Gaussians, and the high-energy continuum peak is modeled as a skew-normal curve. 
We find that the skewness of the main peak at low energies, necessitating the addition of a second Gaussian, is present regardless of the split threshold or grade selection. 
This leads us to believe that it is a real feature of the spectrum and not simply a feature of the detector response.
We believe this is due to the presence of continuum emission which is transmitted through the filter between energies of $1-1.5$ keV (see Fig.~\ref{fig:filters}). 
Even if this is the case, the inclusion of this continuum should not significantly bias our measurement, as the filter transmission drops sharply at lower energies and the H2RG QE in this energy range is fairly uniform (see Fig.~\ref{fig:qe_model}).

\begin{figure}[bt]
    \centering
    \includegraphics[width= 6.5 in]{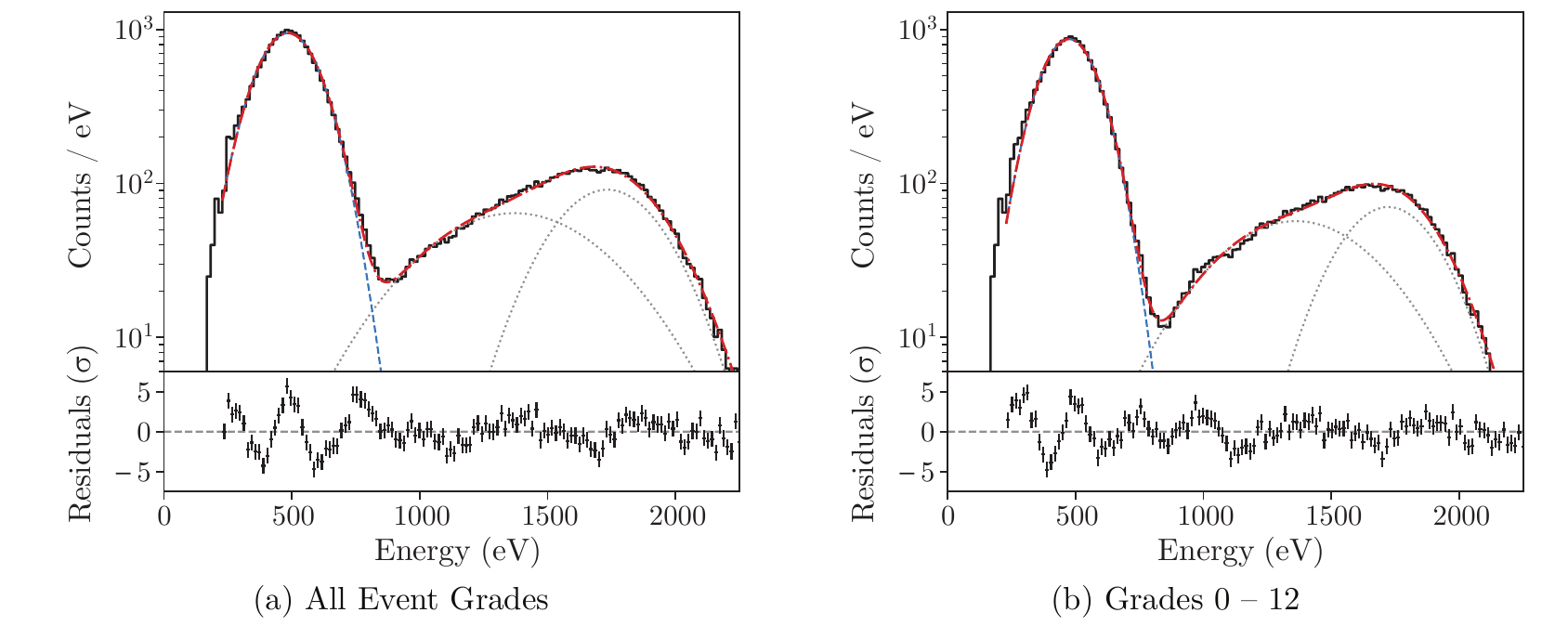}
    \caption{H2RG spectra from the O \ka{} measurement along with the models (red dash-dotted lines) and residuals. Each of the model components is shown, including the Gaussian curve for the characteristic peak (blue dashed line) and the two Gaussians for the continuum/Mg \ka{} peak (gray dotted lines).}
    \label{fig:spec_O}
\end{figure}

Fig.~\ref{fig:spec_O} shows the spectra from the O \ka{} measurement.
The characteristic oxygen peaks are modeled as simple Gaussians. 
For this measurement, we used an oxidized magnesium anode in the Manson source, so the high-energy peak contains a Mg \ka{} line (1.25 keV) in addition to continuum emission.
We modeled this peak as a combination of two Gaussians.

\begin{figure}[bt]
    \centering
    \includegraphics[width= 6.5 in]{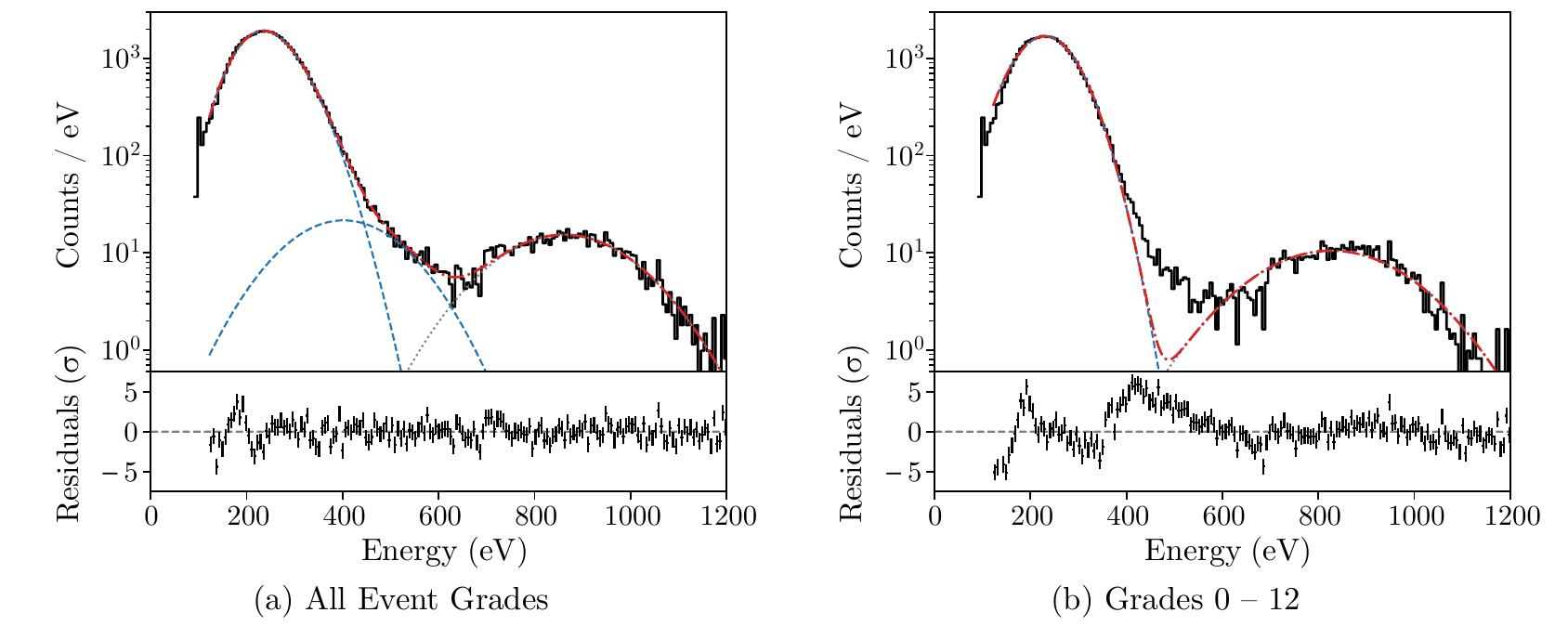}
    \caption{H2RG spectra from the C \ka{} measurement along with the models and residuals. The full models are shown by the red dash-dotted lines, with the included model components shown by blue dashed lines and the excluded components shown by gray dotted lines. Note that we used different models for the C \ka{} peak in the all event grades and grades 0 -- 12 analyses.}
    \label{fig:spec_C}
\end{figure}

Fig.~\ref{fig:spec_C} shows the spectra from the C \ka{} measurement.
In the all-events model, the C \ka{} peak was modeled as a mixture of a skew-normal and normal curve.
This combination was selected in order to sum all C \ka{} events, matching the skewness of the peak caused by the interaction of noise and thresholding effects.
After applying the grades cut, we find that the skewness of the C \ka{} peak is reduced while the high-energy tail is still present.
We ignore the tail in the grades 0 -- 12 analysis, using a single Gaussian to model the C \ka{} peak, in order to focus on only high-quality events whose energy could be accurately determined from a simple analysis. 
Including events in the tail would lead to a $0.4\%$ increase in the count rate and QE in the high-quality event analysis.
In both analyses, the continuum peak is modeled as a Gaussian.

\subsection{QE Calculation}
\label{sec:calc}

The QE is given by the ratio of the flux measured by the H2RG to the flux incident on the detector. 
We determine the absolute flux incident on the PC using the flux measured by the PC and the modeled PC QE (based on window transmission measurements and gas absorption). 
We calculate the flux incident on the H2RG by scaling the incident PC flux by an incident flux ratio ($k$), determined using the Manson beam calibration measurements (described in \S\ref{sec:manson}) or the distances of the detectors from the Fe-55 source.
These factors are combined to obtain the following expression for the H2RG QE:
\begin{equation}
    \text{QE}_\text{\tiny H2RG} = \text{QE}_\text{\tiny PC} \frac{F_\text{\tiny H2RG}}{k \; F_\text{\tiny PC}}.
    \label{eq:qe_calc}
\end{equation}
The values used in this calculation are shown in Table~\ref{tab:QE_calc} for each of the different energies.

\begin{figure}[bt]
    \centering
    \includegraphics[width=0.6 \textwidth]{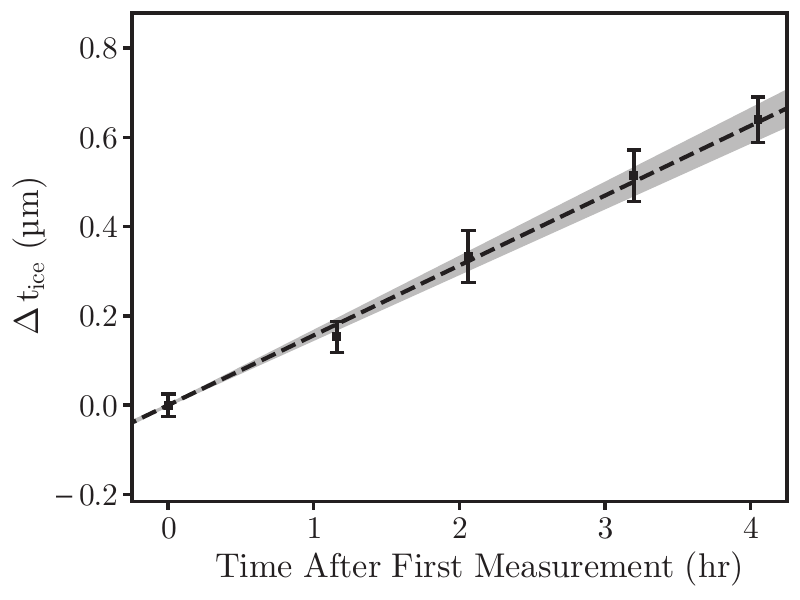}
    \caption{Measurement of the change in ice thickness as a function of time after the first measurement. The data points come from the observed decrease in QE relative to this initial measurement. The dashed line shows the linear fit used to determine the rate of ice accumulation with the uncertainty indicated by the gray shaded region. This rate was used to calculate the thickness of ice on the detector surface during the C \ka{} measurement and correct for attenuation due to this layer.}
    \label{fig:ice_est}
\end{figure}

The C \ka{} analysis is complicated by the presence of an ice layer on the detector during this measurement, due to the lower operating temperature used. 
We correct for the resulting attenuation by measuring the thickness of the ice layer through observations of the decrease in the QE over the course of multiple hours. 
We find that the thickness of the ice layer increased linearly with time, allowing us to infer the thickness of the layer by determining the accumulation rate and the amount of time below the deposition temperature.
Our measurements are shown along with the resulting fit in Fig.~\ref{fig:ice_est}. 
We determine that the thickness of the ice layer was $83\pm14$ nm at the time of our measurement, leading to a transmission of $95.6\pm0.7\%$ for C \ka{} X-rays through the ice layer. 

\begin{table}[bt]
    \caption{Values used for the calculation of the H2RG QE, including the fluxes measured by both detectors, the modeled PC QE, and the ratio of the flux incident on the H2RG to that incident on the PC. Fluxes are given in units of counts s$^{-1}$ cm$^{-2}$. The incident flux ratio for the C \ka{} measurement accounts for the estimated ice attenuation in addition to the measured non-uniformity of the beam.}
    \begin{center}
    \begin{tabular}{cccccccc}
    \toprule
    \textbf{Emission}  & \textbf{PC Flux} & \textbf{PC QE} & \textbf{Incident Flux} & \multicolumn{2}{c}{\textbf{H2RG Flux}}\\
    \textbf{Line}           &                   &                 & \textbf{Ratio}   & \small{Grades 0 -- 12} & \small{All Event Grades} \\
    \midrule
    Mn \ka{}/\kb{} & $13.41 \pm 0.07$ & $69.3 \pm 0.7\%$ & $1.0057 \pm 0.0004$ & $17.91 \pm 0.05$ & $18.40 \pm 0.06$  \\
    Al \ka{}       & $39.42 \pm 0.24$ & $74.9 \pm 1.1\%$ &  $0.991 \pm 0.010$  & $50.46 \pm 0.14$ & $51.30 \pm 0.15$   \\
    O \ka{}        & $19.18 \pm 0.09$ & $38.5 \pm 1.2\%$ &  $1.000 \pm 0.002$  & $35.61 \pm 0.16$ & $42.70 \pm 0.22$    \\
    C \ka{}        & $86.14 \pm 0.36$ & $53.3 \pm 0.8\%$ &  $0.950 \pm 0.008$  & $79.71 \pm 0.56$ & $94.10 \pm 0.39$     \\
    \bottomrule
    \end{tabular}
    \label{tab:QE_calc}
    \end{center}
\end{table}

\subsection{Error Analysis}
\label{sec:error}

We compute the error on our QE measurements using a Monte Carlo analysis, in which we repeatedly sample input variables from their respective distributions in order to sample the QE error distribution.
These input variables include the observed count rates, PC dimensions and gas properties, and incident flux and PC window calibration measurements.
We use $10^5$ simulations, a large enough sample to provide accurate error estimates.
We test the resulting distributions for normality using a Shapiro-Wilk test \cite{Shapiro65} and find that all our QE measurement errors are consistent with Gaussian distributions with P-values greater than $0.05$.  
All uncertainty values provided in this work are the $1\sigma$ values taken from these distributions.

\subsection{QE Results}
\label{sec:results}

\begin{table}[b]
    \caption{Comparison of H2RG QE measurements from the analysis with high-quality grades (Grades 0 -- 12) and with all event grades, shown along the model QE values, which give the fraction of all incident X-ray events expected to be registered by the detector.}
    \begin{center}
    \begin{tabular}{ccccc}
    \toprule
    \textbf{Emission Line} & \textbf{Energy} & \textbf{Measured QE} & \textbf{Measured QE} & \textbf{Model QE} \\
    & \text{(keV)} & \text{(Grades 0 -- 12)} & \text{(All Event Grades)} & {(All X-Ray Events)} \\
    \midrule
    Mn \ka{}/\kb{} & 5.90 / 6.49    &  $92.1 \pm 1.1\%$  & $94.6 \pm 1.1 \%$ & 96.2\% \\
    Al \ka{}     & 1.49             &  $96.7 \pm 1.9\%$  & $98.3 \pm 1.9 \%$ & 97.5\% \\
    O \ka{}      & 0.52             &  $71.3 \pm 2.3\%$  & $85.6 \pm 2.8 \%$ & 89.8\% \\
    C \ka{}      & 0.28             &  $51.9 \pm 1.0\%$  & $61.3 \pm 1.1 \%$ & 62.4\% \\
    \bottomrule
    \end{tabular}
    \label{tab:QE_res}
    \end{center}
\end{table}

\begin{figure}[b]
    \centering
    \includegraphics[width=0.6 \textwidth]{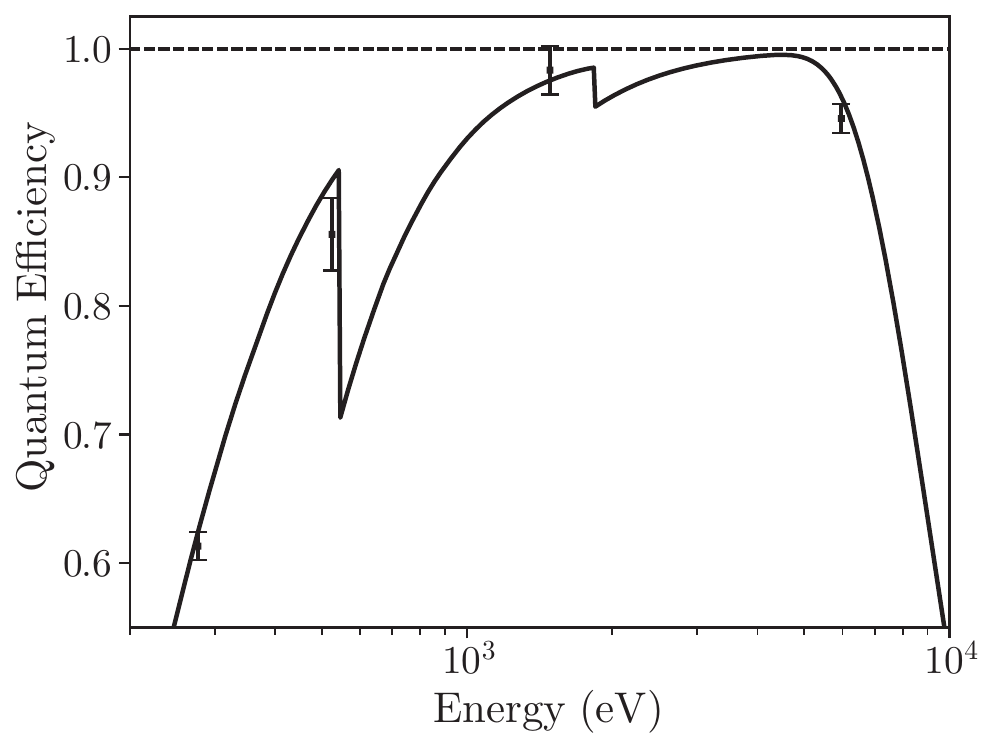}
    \caption{Measured QE values (including all event grades) overplotted with the slab-absorption H2RG QE model (using nominal thicknesses of detector layers). The Fe-55 measurement is plotted at the weighted mean of the Mn \ka{} and \kb{} energies.}
    \label{fig:qe_res}
\end{figure}

Our QE analysis including all event grades yields results which are in good agreement with our simple slab-absorption model.
These results are shown along with the model values (using the nominal values of the absorber and oxide layer thicknesses) in Table~\ref{tab:QE_res} and in Fig.~\ref{fig:qe_res}.
At the Mn \ka{}/\kb{} energies, we measured the effective QE to be $94.6 \pm 1.1\%$.
This value is in good agreement with our modeled effective QE value of $96.2\%$, computed by taking the weighted average of the modeled QE at the Mn \ka{} and \kb{} energies (using the calculated Mn \ka{}/\kb{} ratio specific to our source). 
It is also consistent with the previous measurement made on an H1RG HCD, with a value of $97\pm5\%$ \cite{Bongiorno15}.
This H2RG is expected to have similar QE to H1RGs at moderate energies due to their identical absorber layer thicknesses (100 \microns{}), even as the H1RGs have thinner (250 \AA{}) oxide layers and some devices have deposited aluminum optical blocking layers.

We measured the QE at the Al \ka{} energy to be $98.3 \pm 1.9\%$, also showing excellent agreement with our model and with previous H1RG measurements, which demonstrated $\textrm{QE}\,\gtrsim90\%$ for these detectors between 1 and 3 keV \cite{Prieskorn14}.
As previously noted, a small fraction of continuum X-ray events with energies between 1 and 1.5 keV may be included in this analysis.
This should not significantly affect the analysis, as the QE at these energies should not differ significantly from that at the Al \ka{} energy.
We find that the QE is still high at low energies and remains consistent with our model, with measured values of $85.6\pm2.8\%$ at the O \ka{} energy and $61.3\pm1.1\%$ at the C \ka{} energy.
As the absorption length for these energies is quite small, these measurements indicate that the charge collection efficiency is still high for X-rays absorbed near the surface of the absorber layer.

Our analysis of high-quality events with grades 0 -- 12 still yields high QE, showing only a slight reduction due to the exclusion of events with low-quality grades.
In this analysis, we find an effective QE of $92.1\pm1.1\%$ at the Mn \ka{}/\kb{} energies.
At the Al \ka{} energy, we find a QE of $96.7 \pm 1.9\%$.
At these higher energies, the small difference in QE from the two analyses likely results from events excluded due to excess charge spreading.
The combination of a 15 V substrate voltage, large pixel pitch (36 \microns{}), and high split thresholds makes this a minor effect.
A larger fraction of events are excluded by the grades cut in the lower-energy analyses, as more pixels with high values due to noise are able to trigger the lower split threshold, resulting in events with low-quality grades.
We find a QE of $71.3 \pm 2.3 \%$ at the O \ka{} energy and a value of $51.9\pm1.0\%$ at the C \ka{} energy. 
The next generation of HCDs are expected to have lower read noise and should be able to achieve QE values which are impacted less by grade cuts, as fewer split pixels would be triggered by noise.
These future devices should also have a more Gaussian response at low energies, as thresholds can be better optimized to discriminate pixels to which charge has spread from those with noise.

These measurements represent high QE, comparable to state-of-the-art back-illuminated X-ray CCDs, such as those used in Chandra ACIS \cite{Garmire03}, XMM-Newton EPIC \cite{Hartmann99, Struder01}, or eROSITA \cite{Ebermayer2010, Meidinger21}.
Our H2RG absorber layer was fabricated with a design targeted for optical detectors and was designed with a thicker-than-necessary oxide layer (acting as an optical anti-reflection coating). 
Other X-ray HCDs with thinner oxide layers should theoretically be capable of even higher low-energy QE than that measured on our H2RG.
We anticipate future X-ray HCD models, with lower read noise and absorber layers fabricated explicitly for X-ray applications, to achieve a substantially better low-energy response.
The Lynx High-Definition X-ray Imager \cite{Falcone19} would need high QE, with requirements of $\text{QE}>85\%$ between 0.5 and 10 keV and $\text{QE}>10\%$ between 0.2 and 0.5 keV\cite{Lynx19}.
Our measurements indicate that HCDs are capable of meeting these requirements; however, increasing the thickness of the absorber layer may be necessary to achieve the required QE at the highest energies and decreasing the read noise may be required to achieve higher QE at lower energies when only considering high-quality events.

\begin{figure}[p]
    \centering
    \includegraphics[width=0.6 \textwidth]{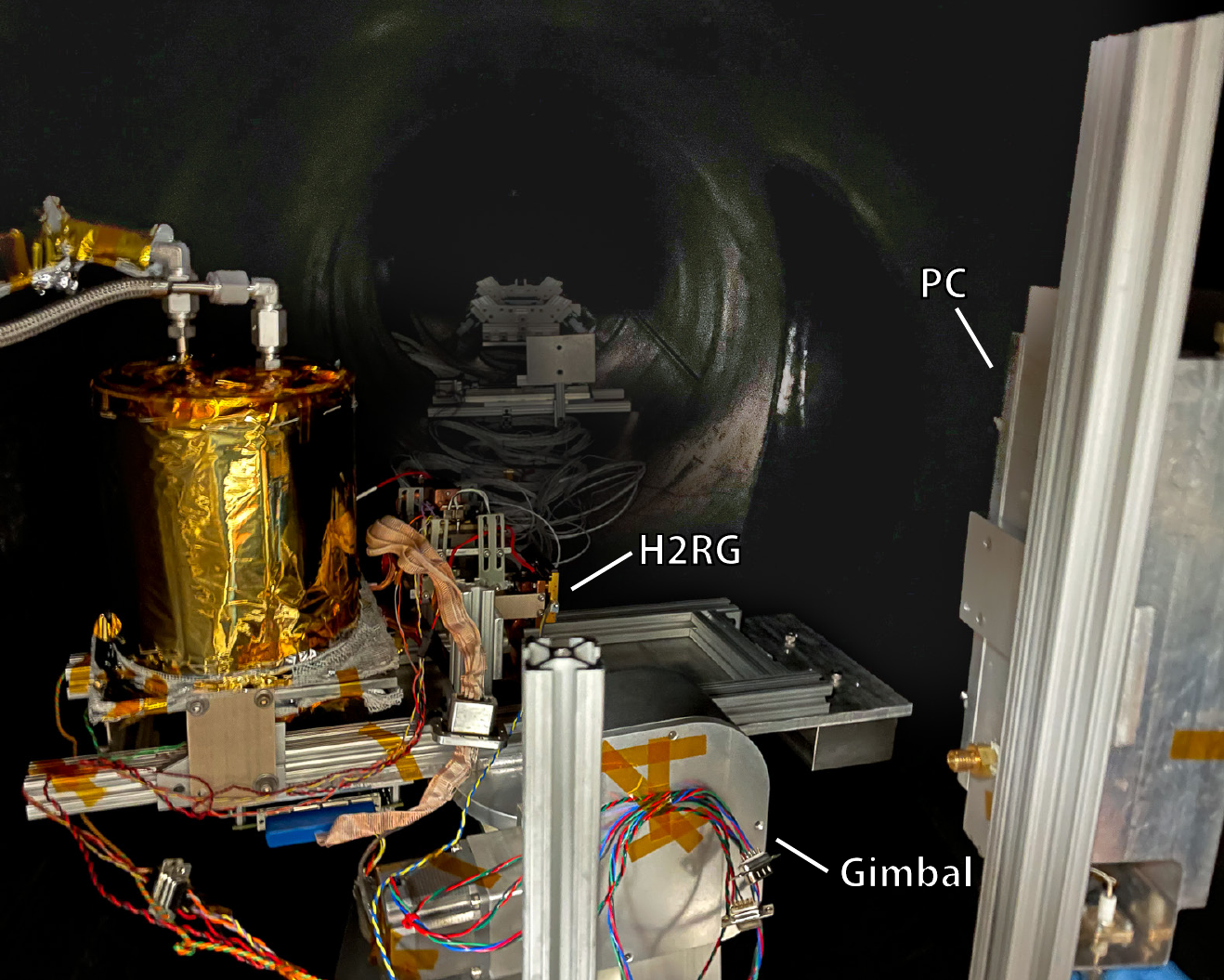}
    \vspace{0.25 cm}
    \caption{Setup of the rotation experiment inside the beamline. The detector can be seen mounted on the gimbal in front of its readout electronics and liquid nitrogen dewar, rotated to an angle of $\sim60^\circ$ off axis. The proportional counter is also visible on the right side of the image.}
    \label{fig:rotation_setup}
\end{figure}

\section{Rotation Measurement}
\label{sec:rotation}

\begin{figure}[p]
    \centering
    \includegraphics[width=0.6 \textwidth]{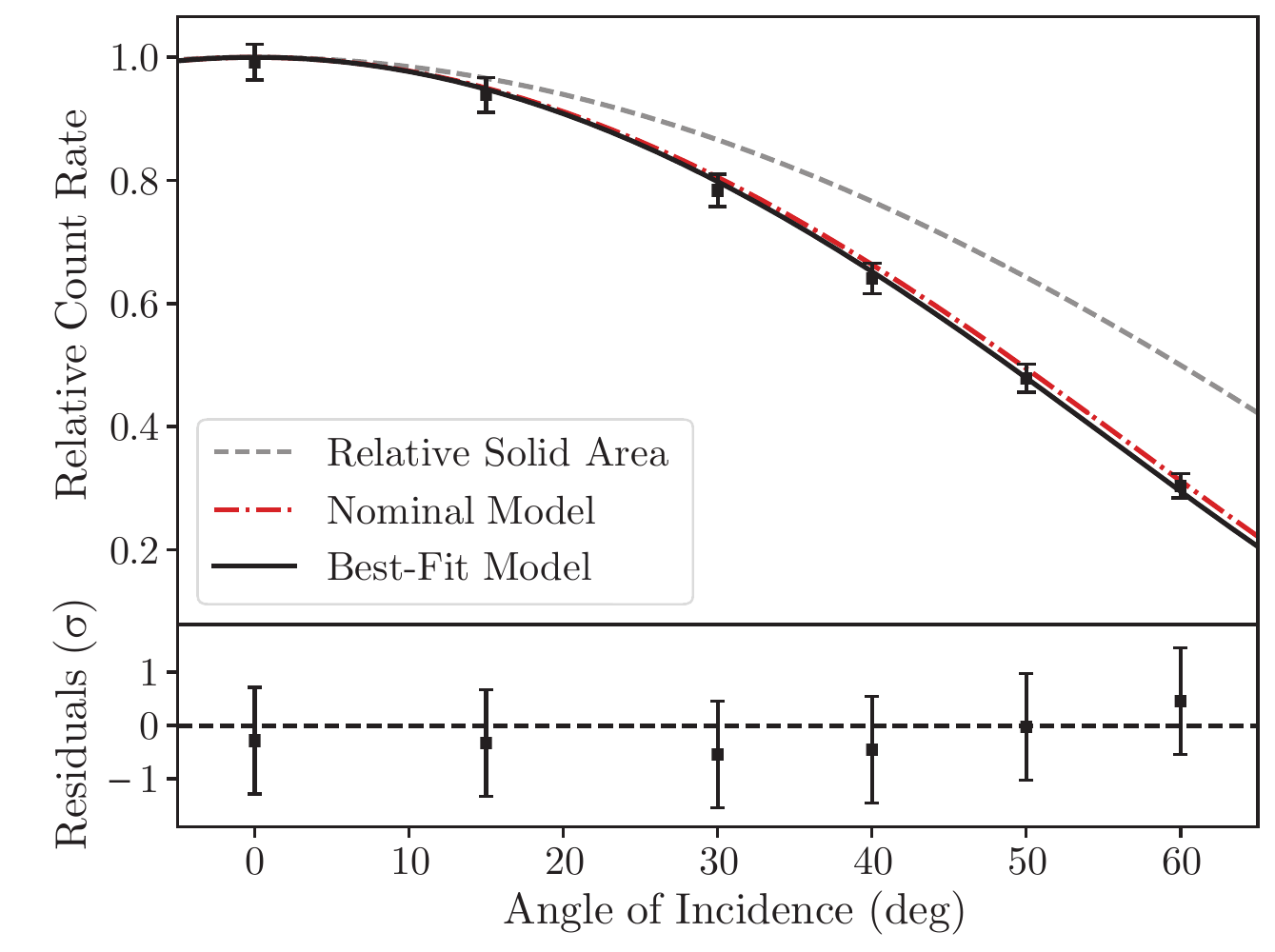}
    \caption{Rotation experiment measurements showing the count rate relative to a prior measurement (face-on at 0$^\circ$), corrected for variation in the Manson flux and ice attenuation. 
    The gray dashed line shows the expected decrease in count rate due to the decrease in solid area exposed to the source. 
    The model for the expected decrease in count rate due to both the decrease in solid area and the increase in path length through the oxide layer is shown for both the nominal (red dash-dotted) and best-fit (solid black) SiO$_2$ layer thicknesses.
    Residuals are shown for the best-fit model.}
    \label{fig:rotation_res}
\end{figure}

As further verification of our low-energy QE measurements and our slab-absorption model, we conducted a rotation experiment to measure the relative QE at the C \ka{} energy as a function of incident angle. 
Changing the incident angle changes the thickness of the oxide layer a photon must traverse to be detected, allowing us to verify the thickness of the oxide layer and probe the detector response to X-rays absorbed at various depths in the detector. 

For the experiment, we mounted the H2RG HCD, internal dewar, and electronics on a rotatable gimbal, shown in Fig.~\ref{fig:rotation_setup}.
We positioned the detector such that its face was centered on the gimbal's axis of rotation, allowing only its angle relative to the source, and not its position, to change between measurements. 
The PC was mounted beside the gimbal and did not rotate with the H2RG assembly. 
It was used as a reference detector to correct for variations in the flux produced by the source. 
We used the same source configuration as was used for the C \ka{} QE measurement, with the Manson source (with the graphite anode) and Al/parylene N filter.

During the measurement, we recorded the count rate of the H2RG as a function of incident angle relative to the source. 
Our analysis for this measurement was similar to that for the all-events carbon analysis, except we fixed parameters of the higher-energy normal component of the C \ka{} model to scale with those of the main skew-normal component, allowing for better fits to the higher-angle measurements with fewer counts.
We also corrected for attenuation due to the ice layer, in a similar manner to the correction described in \S\ref{sec:calc}.

Fig.~\ref{fig:rotation_res} shows the results of the rotation measurement, giving the count rate at each angle relative to an initial face-on measurement, corrected for variation in the source intensity and attenuation by the ice layer. 
Our model accounts for the reduced solid area visible to the source (by a factor of $\cos{(i)}$, where $i$ is the angle of incidence) and greater attenuation as X-rays travel through a larger distance in the oxide layer ($d\propto 1/\cos{(i)}$).
We fit this model to our rotation measurements, allowing the thickness of the oxide layer to vary, and found a best-fit value of $1151 \pm 103$ \AA{}, in reasonable agreement with the nominal value of 1028 \AA{} SiO$_2$.
This best-fit model is shown along with the nominal model  in Fig.~\ref{fig:rotation_res}.
This rotation measurement provides further validation of the slab-absorption model, indicating that it accurately describes charge collection in the H2RG, and provides verification of our low-energy QE results.

\section{Conclusion}
\label{sec:conclusion}

We have measured the QE of a Teledyne Imaging Sensors H2RG HCD at multiple energies using a gas-flow PC as a reference detector. 
Our results show that this detector is capable of achieving high QE across the soft X-ray band, with measured values of $94.6 \pm 1.1 \%$ at the Mn \ka{}/\kb{} energies (5.90/6.49 keV), $98.3 \pm 1.9 \%$ at the Al \ka{} energy (1.49 keV)), $85.6 \pm 2.8 \%$ at the O \ka{} energy (0.52 keV), and $61.3 \pm 1.1 \%$ at the C \ka{} energy (0.28 keV) when including all X-ray events grades. 
These values are consistent with our model, which gives the QE as the fraction of events transmitted by the surface oxide layer that are then absorbed by the absorber layer. 
Our measurements of the relative QE at the C \ka{} energy as a function of angle are also in agreement with this model.
These measurements demonstrate that the detector maintains high charge collection efficiency across the absorber layer and that the QE is not significantly degraded by other factors unaccounted for in the model.

We find the detector still maintains high QE when considering only events with high-quality grades (i.e., grade 0 -- 12 X-ray events, those preferred for studies requiring higher spectral resolution).
In this more restrictive analysis, we measured the QE to be $92.1 \pm 1.1\%$ at the Mn \ka{}/\kb{} energies, $96.7 \pm 1.9\%$ at the Al \ka{} energy, $71.3 \pm 2.3\%$ at the O \ka{} energy, and $51.9 \pm 1.0\%$ at the C \ka{} energy. 
The reduction of QE due to the grades cut is larger at the lower O \ka{} and C \ka{} energies, due to noise artificially triggering split-pixel events.
Future devices with lower read noise should achieve a better low-energy response.
In addition to reducing the instrumental background, this would allow fewer low-energy events to include split pixels triggered by noise, allowing the QE for high-quality events to approach the higher values for all events presented in this work.

HCDs are capable of providing high QE at soft X-ray energies, helping enable future missions to attain the large effective areas necessary to achieve their ambitious science goals.
The faster frame rates enabled by HCDs and other active-pixel sensors will allow thinner optical and UV-blocking filters to sufficiently suppress these backgrounds, thus enabling larger effective areas.
We also anticipate future models of HCDs with lower noise and thinner surface oxide layers.
These factors should further improve the low-energy QE achievable by HCDs, making them excellent candidates for missions which require good response across the soft X-ray band.

\subsection* {Acknowledgments}
This work was supported by a NASA Space Technology Graduate Research Opportunity (grant 80NSSC20K1210), as well as NASA grant 80NSSC20K0778.  
We would also like to acknowledge useful discussions and advice from Yibin Bai at Teledyne Imaging Sensors and Ben Zeiger at Luxel. 
We would also like to thank David Palmer for suggesting the rotation measurement as a method of characterizing detector QE.
In addition, we would like to thank the anonymous reviewers for their helpful comments and suggestions.

%%%%% References %%%%%

\bibliography{report}   % bibliography data in report.bib
\bibliographystyle{spiejour}   % makes bibtex use spiejour.bst

\end{document}